\documentclass[12pt]{article}
\usepackage{graphicx}
\input epsf.sty
\usepackage{mathrsfs}
\usepackage{amsmath}
\usepackage{mathrsfs}
\usepackage{amssymb}
\usepackage{color}
\usepackage{hyperref}
\usepackage{psfrag}

\newcommand{\beq}{\begin{equation}}
\newcommand{\eeq}{\end{equation}}
\newcommand{\Frac}[2]{\frac{\displaystyle #1}{\displaystyle #2}}
\begin{document}
\topmargin 0mm
\oddsidemargin 0mm

\begin{titlepage}
\begin{center}

\hfill USTC-ICTS-12-15\\
\hfill February 2013




\vspace{2.5cm}


{\Large \bf Higgs decays to $\gamma l^+l^-$ in the standard model}

\vspace{10mm}

{Yi Sun$^{1*}$, Hao-Ran Chang$^{2,1\dagger}$, Dao-Neng Gao$^{1\ddagger}$ }
\vspace{4mm}\\{\it\small
 $^1$ Interdisciplinary Center for Theoretical Study,
University of Science and Technology of China, Hefei, Anhui 230026 China}\\
\vspace{4mm}
{\it $^2$ Department of Physics and Institute of Solid State Physics, Sichuan Normal University, Chengdu, Sichuan 610066 China}

\end{center}

\vspace{10mm}
\begin{abstract}
\noindent
The radiative Higgs decays $h\rightarrow\gamma l^+l^-$ with $l = e,~\mu$ and $\tau$ are analyzed in the standard model using $m_h=125$ GeV.  Both tree and one-loop diagrams for the processes are evaluated.  In addition to their decay rates and dilepton invariant mass distributions, we focus on the forward-back asymmetries in these modes. Our calculation shows that the forward-backward asymmetries in $h\to\gamma e^+e^-$ and $h\to\gamma \mu^+\mu^-$ could be up to $10^{-2}$ while in the $\tau^+\tau^-$ final state, these asymmetries are below $1\%$. Thus the forward-backward asymmetries in $h\to\gamma l^+l^-$ might be interesting observables in the future precise experiments both to test our understanding of Higgs physics in the standard model and to probe the novel Higgs dynamics in new physics scenarios.

\end{abstract}
\vfill
\noindent
$^*$ E-mail:~sunyi@mail.ustc.edu.cn\\
\noindent
$^{\dagger}$ E-mail:~hrchang@mail.ustc.edu.cn\\\noindent
$^{\ddagger}$ E-mail:~gaodn@ustc.edu.cn

\end{titlepage}
\section{Introduction}
Since the mass term in the gauge theory violates the gauge invariance, gauge bosons in $SU(2)_L\times U(1)_Y$ gauge field theory should be massless. The Higgs mechanism \cite{higgs} provides a quite simple approach to circumvent this requirement. One can begin with a gauge invariance theory, which has massless gauge bosons and the unstable vacuum, and obtain a theory with massive gauge bosons and the stable vacuum after the algebraic transformation on the Lagrangian due to the spontaneous symmetry breaking of the vacuum.
The simplest form of this mechanism realized in the standard model (SM) requires the existence of a single neutral Higgs boson.
Since all masses of massive particles in the SM are originated from their couplings with Higgs and the Higgs sector plays a key role in our understanding of the nature of the world, the searching for the Higgs boson was one of the most important tasks in the past few decades,  and was also one of the main motivations for the construction of the Large Hadron Collider (LHC) at CERN. After more than forty years' efforts, we finally found a Higgs-like scalar particle at around 125 GeV recently thanks to the hard work of ATLAS \cite{atlas} and CMS \cite{cms} Collaborations at LHC.

Theoretically, the property of the SM Higgs boson has been extensively studied.
The experimental analysis of several different decay channels of the new particle, observed by ATLAS \cite{atlas} and CMS \cite{cms}, shows that its properties are consistent with the elementary Higgs boson in the SM. However, some unexpected signals also appear. The excess events of the diphoton channel $h\to\gamma\gamma$ have drawn many attentions, which have been investigated in a large number of papers using various models \cite{diphoton enhance1, diphton enhance2, diphoton enhance3}. Due to the limitation of the present experimental data, it is still a long way to distinguish these different scenarios.

On the other hand, besides the $h \to \gamma\gamma$ decay, a complementary channel, the radiative Higgs decays $h\to \gamma l^+ l^-$ ($l=e,~\mu$ and $\tau$) could also provide some useful information on the SM Higgs boson. Actually, these radiative decays have been studied in the literature \cite{ABDR97, AR00, only one channel, CQZ12, DR13} and the decay rates and invariant mass distributions have been evaluated. In the last two papers \cite{CQZ12, DR13}, the newly reported mass value of the Higgs boson candidate from \cite{atlas, cms} has been used in their calculations.
The main purpose of the present paper is to focus on the angular distributions, in addition to the rates and invariant mass distributions of the decays. It is known that, in the SM,  $h\to \gamma l^+ l^-$ decays receive the different type contributions from both tree and loop diagrams \cite{ABDR97, AR00, CQZ12}. The angular distribution, due to the interference of these different contributions to the decay amplitudes, will lead to a quantity, called as the forward-backward (FB) asymmetry, which might be an interesting observable in the future precise experiments to further understand the properties of SM Higgs Boson, as well as to explore new physics in the Higgs sector \cite{KK13, LZQ98}.

The paper is organized as follows.
In section 2, the amplitudes of the process $h\rightarrow\gamma l^+l^-$ from the different diagrams are evaluated, and the expressions of the differential decay rate and the FB asymmetries are presented.  Section 3 is our numerical analysis and some discussions on phenomenology. Finally, we summarize our conclusions in section 4, and explicit expressions of some loop functions are given in the Appendix.

\section{The amplitude of $ h \rightarrow \gamma l^+l^-$}\label{section2}
\begin{figure}
  \begin{center} \includegraphics[width=11cm,height=5cm]{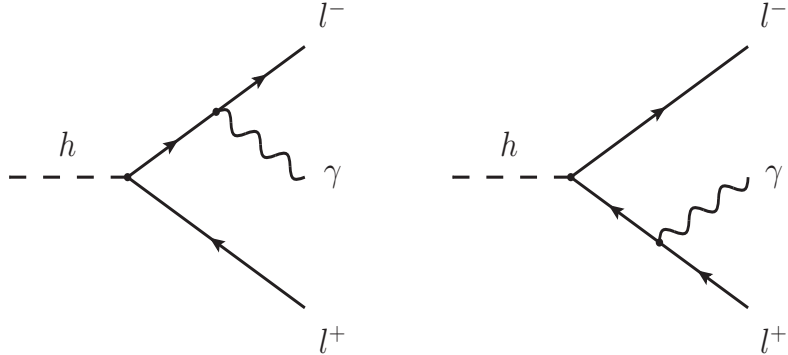}\end{center}
  \caption{The tree diagrams for $h\rightarrow\gamma l^+ l^-$.}\label{tree}
\end{figure}

The tree diagrams of the processes $h\rightarrow\gamma l^+l^-$ are not forbidden, where the photon can be radiated from the lepton leg after Higgs decay to $l^{*}l$, as shown in Figure \ref{tree}. Due to the helicity suppression, this contribution is proportional to the mass of lepton. It will be shown that the contribution of the tree diagrams can be neglected safely in the process $h\rightarrow\gamma e^+e^-$ due to the large mass hierarchy between electron and Higgs, but for $\gamma\mu^+\mu^-$ and $\gamma\tau^+\tau^-$ in the final states, the tree diagrams should be included.
This is consistent with the results in \cite{ABDR97, AR00, CQZ12}.

From Figure \ref{tree}, the decay amplitude of $h\rightarrow\gamma l^+l^-$ at the tree-level can be expressed as follows
\begin{eqnarray}\label{tree level}
i\mathcal{M}_t=i \varepsilon^{\ast}_{\nu}C_0\bar {u}(k_2)\left(\frac{2k_2^{\nu}+\gamma^{\nu}p\!\!\!/}{2k_2\cdot p}-\frac{p\!\!\!/\gamma^{\nu}+2k_1^{\nu}}{2k_1\cdot p}\right)v(k_1),
\end{eqnarray}
 where $C_0=-\frac{2\pi\alpha_e m_l}{m_W\sin\theta_W}$, $m_l$ is the mass of lepton, $\alpha_e$ is the fine-structure constant, $\theta_W$ is the electroweak mixing angle, and $k_1, k_2$ and $p$ represent the momentum of $l^+, l^-$ and $\gamma$ in the final states, respectively.

\begin{figure}
 \begin{center}
  \includegraphics[width=16cm,height=4cm]{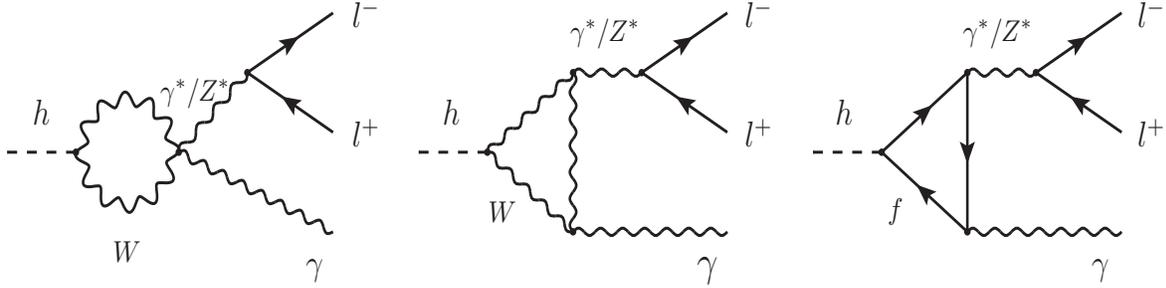}\end{center}
  \caption{The photon and $Z$ pole one-loop diagrams for $h\rightarrow\gamma l^+ l^-$.}\label{three}
\end{figure}

\begin{figure}
\begin{center}
  \includegraphics[width=12cm,height=4cm]{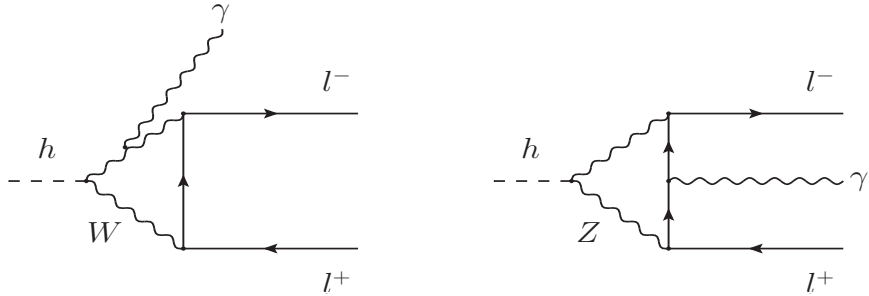}
  \end{center}
  \caption{Four-point box diagrams for $h\rightarrow\gamma l^+l^-$. The photon can also be radiated from another $W$ line in the left diagram and from other charged lepton lines in both diagrams.}\label{fig.1}
\end{figure}

Since its tree-level amplitude is proportional to the mass of lepton, the next-to-leading contributions could be very important for the $h\to\gamma l^+l^-$ transitions, especially for the electron and muon modes. The typical one-loop Feynman diagrams for these processes have been displayed in Figures \ref{three} and \ref{fig.1}, respectively, which are of two basic types: (i) the photon and $Z$ pole diagrams via  $h\rightarrow\gamma \gamma^*\rightarrow\gamma l^+l^-$ and $h\rightarrow\gamma Z^*\rightarrow\gamma l^+l^-$ (Figure \ref{three}); and (ii) four-point box diagrams involving virtual $W$ and $Z$ gauge bosons inside the loop (Figure \ref{fig.1}). It has been pointed out by the authors of Ref. \cite{ABDR97} that type (i) gives the dominant contributions. Our calculations give the consistent conclusion with \cite{ABDR97}. One can check  Figures \ref{differentialwidth}, \ref{differentialwidthmu}, and \ref{differentialwidthtau} in the next section for the numerical analysis, that type (ii) diagrams have only very little contributions. In the remainder of this section, for simplicity, we only show explicitly the results from Figure \ref{three}, to illustrate the discussions on the differential decay rates and the FB asymmetries. Actually, we include the four-point box diagrams' contributions in our numerical calculations of section 3.

The amplitude of $h \rightarrow\gamma l^+l^-$ at the one-loop level  can be expressed as
\begin{eqnarray}
i\mathcal{M}_L&=&i \varepsilon^{\nu\ast}[
\left(p_{\mu }q_{\nu }-g_{\mu \nu }p\cdot q \right)\bar {u}(k_2)(C_1 \gamma^{\mu}+C_2\gamma^{\mu}\gamma^5)v(k_1)\nonumber\\
&&+i \epsilon_{\mu\nu\alpha\beta}p^\alpha q^\beta\bar {u}(k_2)(C_3 \gamma^{\mu}+C_4\gamma^{\mu}\gamma^5)v(k_1)],\label{amplitude}
\end{eqnarray}
where
\begin{eqnarray}
&&C_1=-\left(\frac{1}{4}-\sin^2\theta_W\right)P_Z\Pi_{s\gamma Z}-\frac{1}{q^2}\Pi_{\gamma \gamma},~~~~~~~~~~
C_2=\frac{1}{4}P_Z\Pi_{s\gamma Z},\nonumber\\
&&C_3=-\left(\frac{1}{4}-\sin^2\theta_W\right)P_Z \Pi_{a\gamma Z},~~~~~~~~~~~~~~~~~~~~~~
C_4=\frac{1}{4}P_Z\Pi_{a\gamma Z}\label{ci}
\end{eqnarray}
with
\begin{eqnarray}\label{pz}
P_Z=\frac{1}{\sin\theta_W\cos\theta_W}\frac{1}{q^2-m_Z^2+i m_Z \Gamma_Z}.
\end{eqnarray}
Here we denote $q$ as the momentum of the virtual particle such as $\gamma^*$ or $Z^*$ in Figure \ref{three}, and $q^2=(k_1+k_2)^2$ is lepton pair mass squared. $\Pi_{a\gamma Z}$, $\Pi_{s\gamma Z}$ and $\Pi_{\gamma \gamma}$, induced from  the $Z^*$ and $\gamma^*$ pole diagrams in Figure \ref{three}, respectively, are  given by
\begin{eqnarray}\label{ampofvarious}
\Pi_{a\gamma Z}&=&\frac{\alpha _e^{2}}{m_W\sin\theta_W} \frac{N_c Q_f T_f}{\sin\theta_W\cos \theta_W} A_{f2}\left(\tau _f,\lambda _f\right),\\
\Pi_{s\gamma Z}&=&\frac{\alpha _e^{2}}{m_W\sin\theta_W}\left[-\cot \theta_W A_W\left(\tau_W,\lambda_W\right)\nonumber \right.\\
&&\left.-2 N_c Q_f\frac{ T_f-2Q_f \sin^2\theta_W}{\sin\theta_W\cos\theta_W}A_{f1}\left(\tau _f,\lambda _f\right)\right],\\
\Pi_{\gamma \gamma} &=&\frac{\alpha _e^{2}}{m_W\sin\theta_W}\left[-A_W\left(\tau_W,\lambda_W\right)-4 N_c Q_f^2 A_{f1}\left(\tau _f,\lambda _f\right)\right].
\end{eqnarray}
Here $A_W$, $A_{f1}$, and $A_{f2}$ are functions denoting the contributions from different loops ($W$ loops and fermion loops), with  $\tau_i= \frac{4m_i^2}{m_h^2}, \lambda_i=\frac{4m_i^2}{q^2}(i=f,W)$, and $m_f$ is the mass, $N_c$ is the color multiplicity, $Q_f$, in unit of e, is the charge, $T_f$ is the third component of weak isospin of the fermion $f$ inside the loop. Expressions for loop functions $A_i$'s have been shown in the Appendix.

It is easy to see that the $\gamma^*$ and $Z^*$ pole loop diagrams of Figure \ref{three} are direct extension of the processes $h\rightarrow \gamma\gamma$ \cite{diphoton} and $h\rightarrow \gamma Z$ \cite{photonZ}. One can recover those results of \cite{diphoton, photonZ} by setting $\gamma^*$ and $Z^*$ on-shell in the above expressions excluding the $\Pi_{a\gamma Z}$ parts. The $\Pi_{a\gamma Z}$ part amplitudes, proportional to $A_{f2}$, receive contributions from the fermion loop diagrams in which $Z$ gauge boson is axial-coupling to the loop fermion. This will generate a parity-odd $h\to\gamma Z$ amplitude. As is well-known that the $h\to \gamma Z$ transition is dominated by $W$-loop contribution, which only leads to the parity-even amplitude. The small parity-odd amplitude can be safely neglected in the calculation of $\Gamma(h\to\gamma Z)$ since it does not interfere  with the dominant parity-even amplitude. However, it will be a different story for the $h\to\gamma l^+l^-$ decays because the dominant parity-even $h\gamma Z$ vertex can also generate the parity-odd $h\to\gamma l^+l^-$ amplitude through the axial-coupling of $Z$ gauge boson and leptons in the final states. This means that this parity-odd amplitude cannot be simply abandoned in the present work. Such structures have also been shown in Ref. \cite{CQZ12}. More interestingly, it will be shown below that, these structures ($\Pi_{a\gamma Z}$ part amplitudes) play a key role in the FB asymmetries of the $h\to\gamma l^+l^-$ decays in the SM.

The differential decay rate of $h\to \gamma l^+l^-$, including both tree and loop diagrams contributions, can be written as
\begin{eqnarray}\label{decay rate}
\frac{d\Gamma}{dq^2d\cos\theta}&=&\frac{(m_h^2-q^2)}{512\pi^3m_h^3}
\beta_l [|C_0|^2 A+2Re(C_0 C_1^*)B+2Re(C_0 C_4^*)C\nonumber\\&&+(|C_1|^2+|C_3|^2) D +(|C_2|^2+|C_4|^2) E+2Re(C_1 C_4^*+C_2 C_3^*)F]
\end{eqnarray}
with
\begin{eqnarray}
A&=&\frac{16}{(m_h^2-q^2)^2(1-\beta_l^2\cos\theta)^2}[m_h^4+q^4+32m_l^4-8m_l^2q^2-8m_l^2m_h^2\nonumber\\&&-(m_h^4+ q^4 - 8 m_l^2 q^2 )\beta_l^2 \cos^2\theta],\\ \nonumber\\
B&=&8m_l\frac{m_h^2-q^2+ q^2 \beta_l^2(1-\cos^2\theta)}{1- \beta_l^2 \cos^2\theta},\\
C&=&\frac{8 m_l(m_h^2 -q^2)}{1- \beta_l^2\cos^2\theta}\beta_l\cos\theta,\\
D&=&\frac{(m_h^2-q^2)^2}{2}(q^2+4m_l^2+q^2\beta_l^2\cos^2\theta),\\
E&=&\frac{(m_h^2-q^2)^2}{2}q^2 \beta^2_l(1+\cos^2\theta),\\
F&=&(m_h^2-q^2)^2q^2\beta_l\cos\theta,
\end{eqnarray}
where $\beta_l=\sqrt{1-\frac{4m_l^2}{q^2}}$, $\theta$ is the angle between the three momentum of Higgs boson and the three momentum of $l^-$ in the dilepton rest frame, and the phase space is given by $$4 m_l^2\leq q^2\leq m_h^2,\;\;\;\;-1\leq\cos\theta\leq 1. $$
The contribution of the tree diagrams is given by $A$ term in eq. (\ref{decay rate}) and this term will be divergent when $q^2\rightarrow m_h^2$, which corresponds to the soft photon in the final states. To avoid such divergence, we cut the minimal energy of photon at 1 GeV when we carry out the integration over $q^2$.
$B$ and $C$ terms show the interference between the tree and one-loop diagrams and they are both suppressed by the mass of lepton.
The last three terms $D$, $E$, and $F$, originate from the contribution of the one-loop diagrams.

Although $C$ and $F$ terms, which are proportional to $\cos\theta$, do not contribute to the decay rate, these two terms will lead to a very interesting observable, called as the FB asymmetry, which is defined as
\begin{eqnarray}\label{AFB1}
A_{\rm FB}(q^2)=\Frac{\int_0^1 \frac{d\Gamma }{d q^2d\cos\theta} d\cos\theta-\int_{-1}^0 \frac{d\Gamma }{{dq}^2{d\cos\theta }} d{\cos\theta}}{\int_0^1 \frac{{d\Gamma}}{{d q}^2{d\cos\theta}}  d{\cos\theta}+\int_{-1}^0 \frac{{d\Gamma }}{{dq}^2{d\cos\theta}}  d{\cos\theta}},
\end{eqnarray}
and the corresponding integrated FB asymmetry over $q^2$ is
\begin{eqnarray}\label{AFB2}
\mathcal{A}_{\rm FB}=\Frac{\int_{q^2_{\rm min}}^{q^2_{\rm max}}\int_0^1 \frac{{d\Gamma}}{{dq}^2{d\cos\theta}} d{\cos\theta }dq^2-\int_{q^2_{\rm min}}^{q^2_{\rm max}}\int_{-1}^0 \frac{{d\Gamma}}{{dq}^2{d\cos\theta}} d{\cos\theta}dq^2}{\int_{q^2_{\rm min}}^{q^2_{\rm max}}\int_0^1 \frac{{d\Gamma }}{{dq}^2{d\cos\theta}} d{\cos\theta}dq^2+\int_{q^2_{\rm min}}^{q^2_{\rm max}}\int_{-1}^0 \frac{{d\Gamma}}{{dq}^2{d\cos\theta}}  d{\cos\theta }dq^2},
\end{eqnarray}
where $(q^2_{\rm min}, q^2_{\rm max})$ denotes the range of the integration over $q^2$. From eq. (\ref{decay rate}), one can thus find that the nonzero $C_3$ or $C_4$, which is proportional to $\Pi_{a\gamma Z}$, is required to generate the nonzero FB asymmetries in the SM.  Since $C$ term is proportional to $m_l$, it is expected that $F$ term would give the dominant contributions to the FB asymmetries in $h\to\gamma e^+ e^-$ and $h\to\gamma \mu^+\mu^-$ decays.

We adopt different kinematical variables for the differential decay rate from those in Refs. \cite{CQZ12,DR13} because we are more concerned about the FB asymmetries in the present work. As we shall see, the $\cos\theta$ dependence in the differential decay rate could provide some interesting information in $h\to\gamma l^+l^-$ decays.

\section{Numerical analysis}\label{analysis}

\subsection{Decay rates and invariant mass distributions of $h\to\gamma l^+l^-$}
Both tree and loop diagrams can contribute to the $h\to\gamma l^+l^-$ decays, and the amplitudes at the tree-level are proportional to the mass of the lepton in the final states. Our numerical calculation shows that, for $m_h=125$GeV in the SM,
\beq\label{rate_ee}\frac{\Gamma(h\to\gamma e^+e^-)}{\Gamma(h\to \gamma\gamma)}=5.7\%
\eeq
for the electron mode, and the contribution of tree diagrams is vanishingly small; for the muon mode,
\beq\label{rate_mumu}\frac{\Gamma(h\to\gamma \mu^+\mu^-)}{\Gamma(h\to \gamma\gamma)}=5.8\%,
\eeq
and its tree diagram could play a relevant role, which gives about $30\%$ contribution to the rate; while in the process $h\rightarrow\gamma \tau^+\tau^-$, we have
\beq\label{rate_tautau}\frac{\Gamma(h\to\gamma \tau^+\tau^-)}{\Gamma(h\to \gamma\gamma)}=3.04,
\eeq
where the tree-level contribution is dominant, and the loop diagrams give about $1\%$ contribution.  As mentioned above, in order to avoid the infrared divergence from the tree diagrams, we cut the minimal energy of photon at 1 GeV to illustrate our results of (\ref{rate_ee}), (\ref{rate_mumu}) and (\ref{rate_tautau}).

\begin{figure}
  \begin{center}
  \includegraphics[width=10cm,height=6cm]{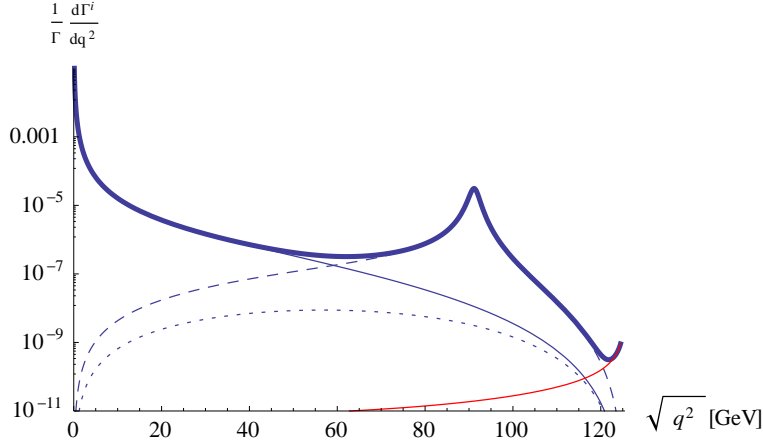}\end{center}
  \caption{The invariant mass distributions of $h\rightarrow \gamma e^+ e^-$ normalized by $\Gamma(h\rightarrow \gamma\gamma)$. The red line denotes the contribution of the tree diagrams, the thin solid line denotes the contribution from the $\gamma^*$ pole diagrams, and the dashed line the contribution from the $Z^*$ pole diagrams while the thick line gives the total contributions. The dotted line denotes the contribution from the four-point box diagrams.}\label{differentialwidth}
\end{figure}
\begin{figure}
  \begin{center}
  \includegraphics[width=10cm,height=6cm]{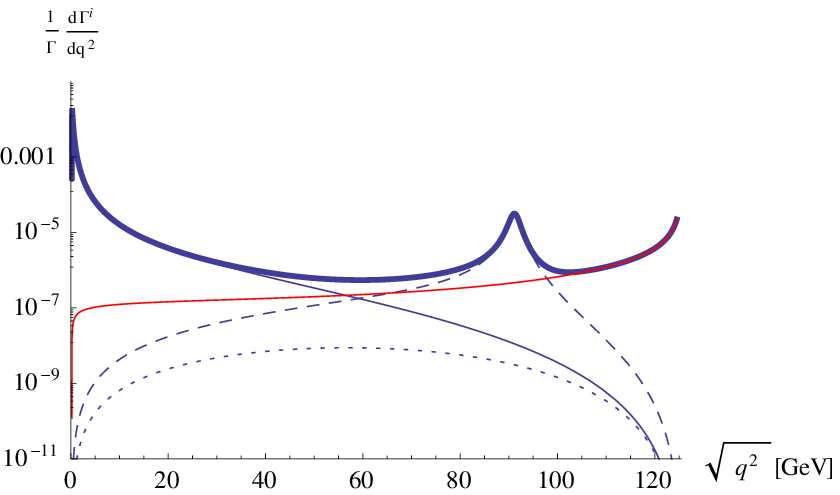}\end{center}
  \caption{The invariant mass distribution of $h\rightarrow \gamma \mu^+ \mu^-$ normalized by $\Gamma(h\rightarrow\gamma\gamma)$. The red line denotes the contribution of the tree diagrams, the thin solid line denotes the contribution from the $\gamma^*$ pole diagrams, and the dashed line the contribution from the  $Z^*$ pole diagrams while the thick line gives the total contributions. The dotted line denotes the contribution from the four-point box diagrams.}
  \label{differentialwidthmu}
\end{figure}

\begin{figure}
  \begin{center}
  \includegraphics[width=10cm,height=6cm]{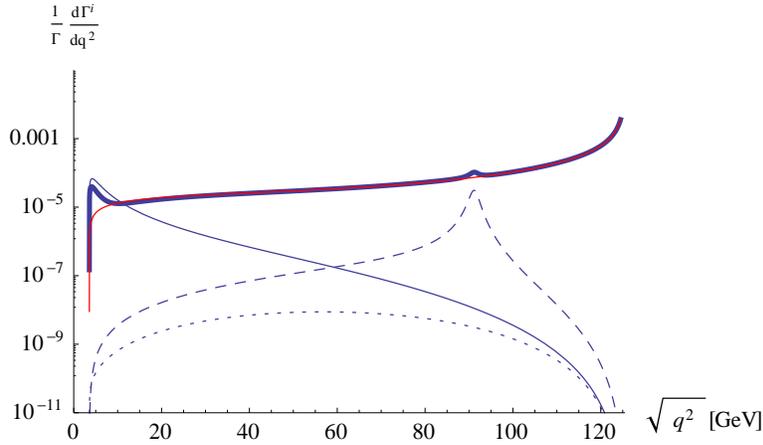}\end{center}
  \caption{The invariant mass distribution of $h\rightarrow \gamma \tau^+ \tau^-$ normalized by $\Gamma(h\rightarrow\gamma\gamma)$. The red line denotes the contribution of the tree diagrams, the thin solid line denotes the contribution from the $\gamma^*$ pole diagrams, and the dashed line the contribution from the  $Z^*$ pole diagrams while the thick line gives the total contributions. The dotted line denotes the contribution from the four-point box diagrams.}
  \label{differentialwidthtau}
\end{figure}

The dilepton invariant mass distributions of $h\rightarrow \gamma l^+l^-$, normalized by $\Gamma(h\rightarrow \gamma\gamma)$, have been given in Figures \ref{differentialwidth}, \ref{differentialwidthmu}, and \ref{differentialwidthtau},  respectively. Different types of contributions including the tree diagrams, $\gamma^*/Z^*$ pole diagrams, and four-point box diagrams, are plotted separately for comparison. It is shown that the four-point box diagrams indeed give a negligible contributions for all three modes, and $\gamma^*/Z^*$ poles (corresponding to the two peaks) give the dominant contributions to the invariant mass distribution in $h\rightarrow \gamma l^+l^-$ with $l=e,~\mu$. From these three plots, one can readily understand our above results of the decay rates, that the tree diagram contribution is vanishingly small in the $e^+e^-$ case, relevant in the $\mu^+\mu^-$ case, and dominant in $\tau^+\tau^-$ channel. Note that these invariant mass distributions in $h\to\gamma e^+ e^-$ and $h\to\gamma\mu^+\mu^-$ have been studied in Ref. \cite{DR13} and Ref. \cite{CQZ12}, respectively. One can find that our plots (Figures 6 and 7) are basically consistent with those results. However, because of the different cuts and different normalization, it is not easy to perform a detailed comparison.

\subsection{Forward-backward asymmetries in $h\to\gamma l^+l^-$}

As expected in section 2, the FB asymmetries $A_{\rm FB}(q^2)$ in $h\to\gamma e^+ e^-$ and $h\to\gamma \mu^+\mu^-$ decays mainly arise from the $F$ term in eq. (\ref{decay rate}), which, by recalling eq. (\ref{ci}), consists of the contributions from the interference between the $\gamma^*$ pole amplitude and the $Z^*$ pole amplitude (referred as $\gamma^*$-$Z^*$ interference), and from the interference among the different amplitudes of $Z^*$ pole diagrams (referred as $Z^*$-$Z^*$ interference); while $C$ term contribution generated from the interference between tree and loop diagrams could be neglected or very small. This is confirmed by our numerical calculations.

\begin{figure}
\begin{center}
  \includegraphics[width=10cm,height=6cm]{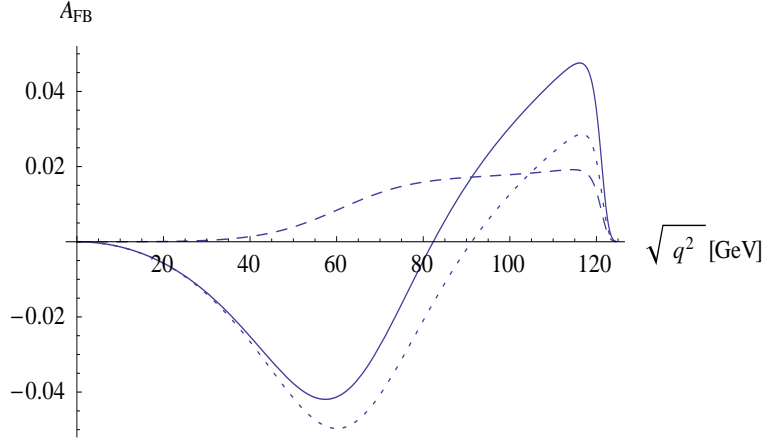}
  \end{center}
  \caption{The FB asymmetry $A_{\rm FB}(q^2)$ in $h\rightarrow \gamma e^+e^-$ as a function of $q^2$. The dotted line denotes the contribution from the $\gamma^*$-$Z^*$ interference, and the dashed line denotes the contribution from the $Z^*$-$Z^*$ interference while the solid line gives the total contribution. The contribution through the interference between the tree diagram and $Z^*$ pole diagrams is extremely small to be shown here.}\label{fbe}
\end{figure}
\begin{figure}
\begin{center}
  \includegraphics[width=10cm,height=6cm]{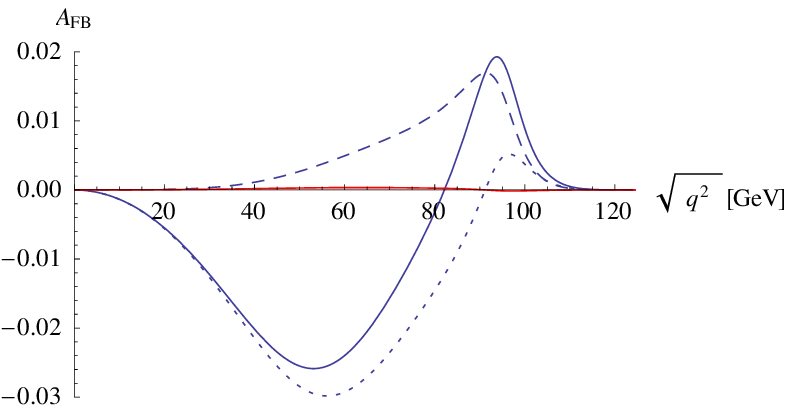}
  \end{center}
  \caption{The FB asymmetry $A_{\rm FB}(q^2)$ in $h\rightarrow \gamma \mu^+\mu^-$ as a function of $q^2$. The dotted line denotes the contribution from the $\gamma^*$-$Z^*$ interference, and the dashed line denotes the contribution from the $Z^*$-$Z^*$ interference, and the red line denotes the very small contribution through the interference between the tree diagram and $Z^*$ pole diagrams while the solid line gives the total contribution.}\label{mufbe}
\end{figure}
\begin{figure}
\begin{center}
  \includegraphics[width=10cm,height=6cm]{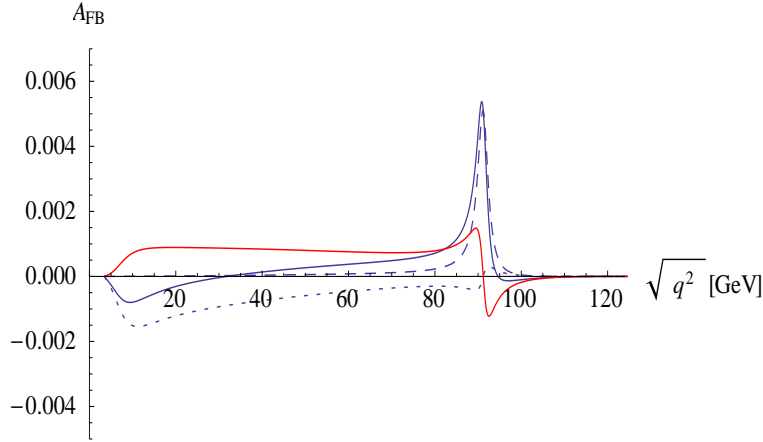}
  \end{center}
  \caption{The FB asymmetry $A_{\rm FB}(q^2)$ in $h\rightarrow \gamma \tau^+\tau^-$ as a function of $q^2$. The dotted line denotes the contribution from the $\gamma^*$-$Z^*$ interference, the dashed line denotes the contribution from the $Z^*$-$Z^*$ interference, and the red line denotes the contribution through the interference between the tree diagram and $Z^*$ pole diagrams while the solid line gives the total contribution.}\label{taufbe}
\end{figure}

The FB asymmetries $A_{\rm FB}(q^2)$ in $h\rightarrow \gamma e^+e^-$ and $h\rightarrow \gamma \mu^+\mu^-$ have been plotted in Figures \ref{fbe} and \ref{mufbe}, respectively, which, in most kinematical region, could be up to $10^{-2}$. One can see that the $C$ term contribution is very small for the muon mode,  and we even do not plot this part contribution in Figure \ref{fbe} since it is extremely small to be shown.   Figure \ref{taufbe} gives the FB asymmetry in $h\to\gamma\tau^+\tau^-$, which is about one order of magnitude smaller than the former two modes. It is thought that $C$ term contribution in eq. (\ref{decay rate}) might be important for $\tau$ channel, however, the tree-level amplitude is strongly dominant over the other amplitudes, which, from eq. (\ref{AFB1}), obviously leads to the suppression of $A_{\rm FB}(q^2)$ in $h\to\gamma\tau^+\tau^-$.
\begin{table}
\centering
\begin{tabular}{|c|c|c|c|c|c|c|c|c|c|}
  \hline
  $q^2_{\rm min}$ - $q^2_{\rm max}$(GeV$^2$)& 10$^2$-$30^2$ & 30$^2$-$50^2$ & 50$^2$-70$^2$ & 70$^2$-90$^2$ & 90$^2$-110$^2$ & full phase space\\
  \hline$h\to\gamma e^+e^-$ & -0.5\% & -2.3\% & -3.7\% & 0.8\% & 1.9\%& 0.4\% \\
  \hline$h\to\gamma \mu^+\mu^-$ & -0.4\% & -1.9\% & -2.3\% & 0.7\% & 1.5\%& 0.4\%\\
  \hline
\end{tabular}\caption{The integrated FB asymmetries $\mathcal{A}_{\rm FB}$ for some cuts on $q^2$ in $h\rightarrow \gamma e^+e^-$ and $h\rightarrow \gamma \mu^+\mu^-$ decays.}\label{mu}\end{table}

As shown in Figures \ref{fbe} and \ref{mufbe}, $A_{\rm FB}(q^2)$ in $h\to \gamma e^+e^-$ and $h\to\gamma\mu^+\mu^-$ will be from negative to positive when $q^2$ is running in the phase space. This comes from that the $\gamma^*$-$Z^*$ interference will change sign for $q^2$ crossing $m_Z^2$, since, in this case, the real part of  $P_Z$ in eq. (\ref{pz}) changes sign. Thus the integrated FB asymmetries $\mathcal{A}_{\rm FB}$  defined in eq. (\ref{AFB2}) may be not very significant if we integrate $q^2$ in the full phase space. On the other hand, when we calculate these integrated asymmetries for some cuts on $q^2$, $\mathcal{A}_{\rm FB}$ could be at percent level. The corresponding numerical results have been given in
Table \ref{mu}, and $\mathcal{A}_{\rm FB}$ in the full phase space of $q^2$ is only $0.4\%$ for both modes. While for some kinematical regions, these asymmetries might be enhanced, for instance,  $50 ~\text{GeV}<\sqrt{q^2}<70 ~\text{GeV}$, the magnitude of $\mathcal{A}_{\rm FB}$ is $3.7\%$ for $h\rightarrow \gamma e^+e^-$ and is $2.3\%$ for $h\rightarrow \gamma \mu^+\mu^-$. In the process $h\to\gamma \tau^+\tau^-$, the integrated FB asymmetry will be also very small,  $\mathcal{A}_{\rm FB}$$\sim 0.02\%$ in the full phase space of $q^2$. Therefore it is not significant to calculate them for any cuts on $q^2$.

\section{Summary and Conclusions}\label{conclusion}

We have presented a comprehensive analysis of the radiative Higgs decays $h\to \gamma l^+l^-$ with $l=e,~\mu$ and $\tau$ in the SM, by using $m_h$=125 GeV, the newly reported mass value of the Higgs boson candidate from LHC \cite{atlas, cms}. Both tree and one-loop Feynman diagrams for the processes are evaluated. It is found that loop amplitudes are dominant for both electron and muon modes. The tree-level contribution can be fully neglected in $h\to\gamma e^+ e^-$ decay, and for the muon mode, it will give a sizable contribution to the decay rate while the tree-level transition is very important in $h\to \gamma \tau^+\tau^-$. Our numerical results show that $\Gamma(h\to\gamma l^+l^-)$ with $l=e,~\mu$ is about $6\%$ of $\Gamma(h\to\gamma\gamma)$, and $\Gamma(h\to\gamma\tau^+\tau^-)$ is about factor three of the diphoton rate, which is almost entirely contributed by the tree diagrams.
The dilepton invariant mass distributions, normalized by $\Gamma(h\to\gamma\gamma)$, have been analyzed for all three modes. Different types of contributions are plotted separately for comparison, the four-point box diagrams are found to be negligible, and the photon and $Z$ pole diagrams give the dominant contribution in $h\to\gamma e^+e^-$ and $h\to\gamma \mu^+\mu^-$. Our results are basically consistent with those of Ref. \cite{CQZ12} and Ref. \cite{DR13}, in which the different cuts and normalization have been used.

One of the main motivation of the present work is to investigate the FB asymmetries in $h\to\gamma l^+l^-$. Different from the two-body decay $h\to \gamma\gamma$, three-body radiative decay may generate a nontrivial angular distribution, which would provide some complementary information for the diphoton decay. In the SM, thanks to the parity-odd $h\to\gamma Z^*$ amplitude induced from the fermion loop, nonzero FB asymmetries can be expected, and we have explicitly evaluated the SM contributions to $A_{\rm FB}(q^2)$ and ${\cal A}_{\rm FB}$ in these decays. Interestingly, these asymmetries can be up to $10^{-2}$ in $h\to\gamma e^+e^-$ and $h\to\gamma \mu^+\mu^-$ decays while they are suppressed below $1\%$ for the $\gamma \tau^+\tau^-$ final states.
Consider the magnitude of these asymmetries, the measurement of them is indeed not an easy task; nevertheless, experimental studies of the FB asymmetries in $h\to \gamma l^+l^-$ decays in the future would be very helpful both to increase our understanding of the properties of SM Higgs boson and to probe the novel couplings of Higgs in the new physics scenarios.

In summary, our present analysis indicates that the radiative Higgs decays $h\to\gamma l^+l^-$ are worth to be seriously considered in the future experimental investigations. These decays, due to their complementarity to the $h\to\gamma\gamma$ decay,  may provide us with some valuable information on the Higgs physics in the SM and its possible extensions.

\section*{Acknowledgements}

This work was supported in part by the NSF of China under Grant Nos. 11075149 and 11235010, and the 973 project under grant No. 2009CB825200.

\appendix
\newcounter{pla}
\renewcommand{\thesection}{\Alph{pla}}
\renewcommand{\theequation}{\Alph{pla}.\arabic{equation}}
\setcounter{pla}{1}
\setcounter{equation}{0}

\section*{Appendix: Loop functions}

Here we explicitly show the expressions for some loop functions in section 2.
\begin{eqnarray}
A_{f1}(\tau,\lambda)&=&I_1(\tau,\lambda)-I_2(\tau,\lambda)\,\,\,,\\
A_{f2}(\tau,\lambda)&=&\frac{\tau\lambda}{\lambda -\tau }[2 g(\tau)-2 g(\lambda) + f(\tau) - f(\lambda)]\,\,\,,\\
A_W(\tau,\lambda)&=&\left[\left(1+\frac{2}{\tau}\right)\left(\frac{4}{\lambda}-1\right)-\left(5+\frac{2}{\tau}\right)\right]
I_1(\tau,\lambda)\nonumber \\&& +16\left(1-\frac{1}{\lambda}\right)I_2(\tau,\lambda)
\end{eqnarray}
with
\begin{eqnarray}
I_1(\tau,\lambda)&=&\frac{\tau\lambda}{2(\tau-\lambda)}+
\frac{\tau^2\lambda^2}{2(\tau-\lambda)^2}[f(\tau)-f(\lambda)]+
\frac{\tau^2\lambda}{(\tau-\lambda)^2}[g(\tau)-g(\lambda)]\,\,\,,\\
I_2(\tau,\lambda)&=&-\frac{\tau\lambda}{2(\tau-\lambda)}[f(\tau)-f(\lambda)]\,\,\,,
\end{eqnarray}
where
\begin{equation}
f(\tau)=\left\{
\begin{array}{lr}
\arcsin^2\frac{1}{\sqrt{\tau}}&\tau\ge 1\\
-\frac{1}{4}\left[\ln\frac{1+\sqrt{1-\tau}}{1-\sqrt{1-\tau}}-i\pi\right]^2&
\tau<1\,\,\,,
\end{array}\right.
\end{equation}
\begin{equation}
g(\tau)=\left\{
\begin{array}{lr}
\sqrt{\tau-1}\arcsin\frac{1}{\sqrt{\tau}}&\tau\ge 1\\
\frac{\sqrt{1-\tau}}{2}\left[\ln\frac{1+\sqrt{1-\tau}}{1-\sqrt{1-\tau}}-i\pi\right]&
\tau<1.
\end{array}\right.
\end{equation}
For $m_h=125$ GeV, $\tau<1$ will not be used if we only consider $W$ and top quark loop.

\end{document}